\documentclass{iau}
\usepackage{amsmath}
\usepackage{graphicx}
\usepackage{multirow}

\begin{document} 

\lefttitle{Sergei I. Ipatov}
\righttitle{Migration of ejected bodies}

\jnlPage{1}{5}
\jnlDoiYr{2024}
\doival{10.1017/xxxxx}

\aopheadtitle{Proceedings IAU Symposium No. 393}
\editors{Youssef Moulane, ed.}

\title{Migration of bodies ejected from the Earth and the Moon}%

\author{Sergei I. Ipatov}
\affiliation{Vernadsky Institute of Geochemistry and Analytical Chemistry of RAS, 
\\ 119991, 19 Kosygin st., Moscow, Russia \\ email: {\tt siipatov@hotmail.com}}


\begin{abstract}
This study analyzes the motion of bodies ejected from the Earth or the Moon. We studied the ejection of bodies from several points on the Earth's surface, as well as from the most far point of the Moon from the Sun. Different velocities and angles of ejection of bodies were considered. The dynamical lifetimes of bodies reached a few hundred million years. Over the entire considered time interval, the values of the probability of a collision of a body ejected from the Earth with the Earth were approximately 0.3, 0.2, and 0.15–0.2 at an ejection velocity $v_{ej}$ equaled to 11.5, 12, and 14 km/s, respectively. At $v_{ej}$$\le$11.3 km/s, most of the ejected bodies fell back onto the Earth. 
The total number of bodies delivered to the Earth and Venus probably did not differ much. The probabilities of collisions of bodies with Mercury and Mars usually did not exceed 0.1 and 0.02, respectively. 
At $v_{ej}$$ \ge$11.5 km/s, the probability of a collision of a body ejected from the Earth  with the Moon was about 15–35 times less than that with the Earth, and it was about 0.01. The probability of a collision with the Earth for a body ejected from the Moon moving in its present orbit was about 0.3–0.32,   0.2–0.22, and 0.1–0.14 
at $v_{ej}$=2.5 km/s, $v_{ej}$=5 km/s,  and at 12$\le$$v_{ej}$$\le$16.4 km/s, respectively. 
\end{abstract}

\maketitle

\section{Introduction}
The terrestrial planets and the Moon were bombarded by impactors during their formation and at the stage of the Late-Heavy Bombardment. 
Some material could be ejected at such impacts.
Earlier \cite{Gladman2005} and \cite{Reyes-Ruizetal2012} studied the motion of bodies ejected from the Earth  during time interval equal to 30 Kyr. Considered initial velocities were perpendicular to the surface of the Earth. According to (\cite{Shuvalov2011}), the values of an ejection angle $i_{ej}$ are mainly between $20^\circ$ and $55^\circ$, especially between $40^\circ$ and $50^\circ$. 
More detailed introduction information on the item of the below paper can be found in
(\cite{Ipatov2024}, \cite{Ipatov2025}).

\section{The model and initial data for studies of ejection of bodies from the Earth}

In each calculation variant, the motion of 250 ejected bodies was studied for the same ejection point on the Earth's surface and for the fixed values of an ejection angle $i_{ej}$, a velocity $v_{ej}$ of ejection, and a time step $t_s$ of integration. In different variants, the values of $i_{ej}$ equaled to 15$^\circ$, 30$^\circ$, 45$^\circ$, 60$^\circ$, 89$^\circ$, or 90$^\circ$, and $v_{ej}$ equaled mainly to 11.22, 11.5, 12, 12.7, 14, 16.4, or 20 km/s. In most calculations, bodies started directly from the Earth. In each variant, bodies started from one of six considered opposite points on the Earth's surface (\cite{Ipatov2024}). Bodies ejected from points F and C started their motion from the most far and most close points to the Sun, respectively. For points W and B, bodies started from the points on Earth's surface located on the forward side (in the direction of motion) and on the opposite side, respectively. For points U and D, bodies started from the Earth at maximum and minimum values of z at Oz axis perpendicular to the plane of the orbit of the Earth. The rotation of the Earth around its axis was not taken into account, as the additional velocity of its surface caused by such rotation is smaller than the considered ejection velocity. The velocity of rotation doesn't exceed 0.5 km/s, and it is different for different points on the Earth's surface.

The motion of bodies was studied during the dynamical lifetime $T_{end}$ of all bodies, which equaled to a few hundreds of Myr and could exceed 1 Gyr. During this time interval, all bodies either collided with planets or the Sun, or they were ejected into hyperbolic orbits. The gravitational influence of the Sun and all eight planets was taken into account. Bodies that collided with planets or the Sun or reached 2000 AU from the Sun were excluded from integration. 
The motion of the Moon was not included in the integration of motion equations. The probabilities of collisions of bodies with the Moon were calculated with the use of some formulas. Such calculations were based on the arrays of orbital elements of the bodies that left the Hill sphere of the Earth (\cite{Ipatov2024}). After a body had left the Hill sphere, it could enter this sphere again several times during its long dynamical lifetime. For such body, the contribution to the probability of its collision with the Moon was relatively small before the body left the Hill sphere for the first time. However, if an ejected body moved for a long time inside the Hill sphere of the Earth, then the probability of its collision with the Moon could be not small. Actually such long motion inside the Hill sphere could be only for a small ejected velocity especially if the Moon embryo  moved close to the Earth (see Section 6) or if the ejected material was not in the form of solid bodies (e.g. was in the form of vapor).

The symplectic code from the SWIFT integration package  by \cite{Levison1994}
was used for integration of the motion equations. In different variants the considered time integration step $t_s$ equaled to 1, 2, 5, or 10 days. The results of calculations with different $t_s$ were compared. Most of calculations have been made with an integration time step equal to 5 days. With this step, the obtained probabilities of collisions with planets were about the same as those obtained with a smaller $t_s$.

\section{Probabilities of collisions of bodies ejected from the Earth with the Earth}

Exclusive for ejection velocities close to the parabolic velocity, on average, about a half and less than 10\% of initial ejected bodies still moved in elliptical orbits after 10 and 100 Myr, respectively. A few ejected bodies collided with planets after 250 Myr.  Impactors with a diameter of about 10 km formed Chicxulub and Popigai craters on the Earth with ages 66 and 36 Myr. About one 1-km impactor could collide the Earth during 1 Myr. Some bodies ejected from the Earth during the last 100 Myr can still move in the zone of the terrestrial planets and can collided with them.

The probability $p_E$ of collisions of bodies ejected from the Earth with the Earth is greater for a smaller ejection velocity. Most of the ejected bodies fell back onto the Earth at $v_{ej}$ slightly greater than the parabolic velocity (at $v_{ej}$$\le$11.25 km/s). At $v_{ej}$=11.5 km/s, the values of $p_E$ were about 0.2-0.3 and did not differ much for different starting points on the Earth's surface. At $v_{ej}$=12 km/s, $p_E$ was about 0.16-0.27.  For the point W (when bodies were ejected from the forward point of the Earth's motion - from apex), the probabilities $p_E$ can differ much from those for other points. For this point at $v_{ej}$$\ge$16.4 km/s and 
$45^\circ$$\le$$i_{ej}$$\le$$60^\circ$, the value of $p_E$ could be 0. In this case more than 80\% of the bodies were thrown into hyperbolic orbits, and most of other bodies collided with the Sun. 
At $v_{ej}$=20 km/s, the values of  $p_E$ equaled to 0  for ejection from the forward point W at  $i_{ej}$$\ge$$30^\circ$. They exceeded 0.2 for ejection from the back point B (from antapex), and were about 0.1 for ejection from other four considered points of ejection. 
The ratios of the probability of collisions of bodies with the Earth to the probabilities of collisions of bodies with other planets and the Sun usually decreased with time. 

Time variations of the probabilities of collisions of bodies ejected from the Earth with the Earth and the Moon were considered by
\cite{Ipatov2024} for time interval $T$ between 10 Kyr and 100 Myr. The probabilities (especially at smaller $T$) depended much on $v_{ej}$   and the point of ejection. 
At $i_{ej}$$=$$45^\circ$ and $v_{ej}$=12 km/s for all 6 ejection points,  the ratio of the values of $p_E$ at $T$ to those at $T_{end}$ was in the ranges      
0.02-0.04,  0.2-0.3, and  0.6-0.8 at $T$ equal to 10 Kyr, 1 Myr, and 10 Myr, respectively. The intervals of the values of $p_E$ for the above case were  0.004-0.008, 0.06-0.07, and 0.15-0.16 at  $T$ equal to 10 Kyr, 1 Myr, and 10 Myr, respectively.  The above intervals can be wider for $v_{ej}$$\ge$12.7 km/s.   

The average velocities of collisions of  bodies ejected from the Earth with the Earth and the Moon are greater for greater
 ejection velocity. The collision velocities of bodies with the Earth were about 13, 14–15, 14–16, 14–20, 14–25, and 15–35 km/s at $v_{ej}$ equal to 11.3, 11.5, 12, 14, 16.4, and 20 km/s, respectively.
The collision velocities of ejected bodies with the Moon were mainly in the range of 7–8, 10–12, 10–16, 11–23, and 12–33 km/s at $v_{ej}$ equal to 11.3, 12, 14, 16.4, and 20 km/s, respectively.

\section{Probabilities of collisions of bodies ejected from the Earth with the Moon}

The probability $p_M$=$p_E/k_{EM}$  of collisions of bodies ejected from the Earth with the Moon was calculated to be equal to the ratio of the value of $p_E$ obtained at integration to the ratio $k_{EM}$ of the probabilities of collisions of bodies with the Earth and the Moon. The value of  $k_{EM}$ was calculated with the use of some formulas based on the arrays of orbits of bodies during considered time interval $T$   similar to (\cite{Ipatov2019}). 
At $v_{ej}$$ \ge$11.5 km/s, the probability of a collision of a body ejected from the Earth  with the Moon was about 15–35 times less than that with the Earth.
 For the Moon moved in its present orbit, the probability $p_M$  was of the order of 0.01. 
It was about 0.01-0.02 at  $v_{ej}$=11.3 km/s and 0.004-0.008 at  $v_{ej}$$\ge$14 km/s.
Based on this result, \cite{Ipatov2024} concluded that a large Moon embryo should be formed close to the Earth in order to accumulate material not rich in iron ejected from forming Earth. The estimates of $p_M$ have been made for the model of ejection of bodies. 
In the multi-impact and giant-impact models, some of the ejected matter (e.g. in the form of vapor) went into orbits around the Earth. In this case the growth of the Moon embryo could be greater than for the ejection of bodies considered above.
 \cite{Ipatov2018} supposed that the embryos of the Earth and the Moon with masses smaller by one or two orders of magnitude than their present masses could be formed as a result of contraction of a common rarefied condensation. 

\section{Probabilities of collisions of bodies ejected from the Earth with planets and probabilities of their ejection into hyperbolic orbits}

The probability $p_V$ of collisions of the bodies ejected from the Earth with Venus during a whole considered time interval of evolution (at $T$=$T_{end}$) were often about 0.2-0.35 at 11.5$\le$$ v_{ej} $$\le$16.4 km/s. The values of $p_V$ were were less than $p_E$ at $v_{ej}$$\le$11.4 km/s, were about $p_E$ at $v_{ej}$=11.5 km/s, and were greater than $p_E$ at $v_{ej}$$ \ge$12 km/s. The total number of ejected bodies delivered to the Earth and Venus probably did not differ much, and the upper layers of Earth and Venus can contain similar material.

The probability of collisions of bodies ejected from the Earth  with Mercury during a whole considered time interval was about 0.02-0.08 and 0.03-0.05 at 11.3$\le$$v_{ej}$$\le$11.5 and 12$\le$$v_{ej}$$ \le$20 km/s, respectively. The probabilities of collisions of bodies ejected from the Earth with Mars were smaller than those with Mercury. At $T$=10 Myr and $T$=$T_{end}$,  the probability of collisions of bodies with Mars did not exceed 0.012 and 0.025 , respectively. Therefore, more material ejected from the Earth was delivered to Mercury than to Mars. The fraction of bodies ejected from the Earth that collided with Mars in 1 Myr was about 0.0002. It is supposed by
\cite{Mileikowsky et. al (2000)}
 that microbes can survive during the flight between the Earth and Mars with duration of about 1 Myr.

The fraction $p_J$ of bodies ejected from the Earth and then collided with Jupiter was of the order of 0.001. 
However, at $T$=$T_{end}$ in a few variants, mainly with large $v_{ej}$, the value of $p_J$ exceeded 0.01.
The probability of collisions of bodies ejected from the Earth with the Sun often was between 0.2 and 0.5 at $v_{ej}$$ \ge $11.3 km/s.


The fraction of bodies that had been ejected from the Earth and then were ejected into hyperbolic orbits
did not exceed 0.1 at $v_{ej}$$\le$12 km/s and $T$=$T_{end}$.
It was mainly greater for a greater ejection velocity. 
 At  $v_{ej}$$\ge$16.4 km/s and $i_{ej}$$ \ge$60$^\circ$, all bodies were ejected from the Solar System if they started from the front point W. 
The fraction of bodies ejected into hyperbolic orbits was about 0.25-0.27 at  $i_{ej}$=45$^\circ$ and $v_{ej}$=16.4 km/s 
if points of ejection were not in the front or back of the Earth's motion.    
More detailed studies of the probabilities of bodies ejected from the Earth with the Moon were discussed by \cite{Ipatov2024}, and such probabilities with planets were presented by \cite{Ipatov2025}.

\section{Probabilities of collisions of bodies ejected from the Moon with the Earth} 

Migration of bodies ejected from the Moon was studied similar to the ejection of bodies from the Earth from  point F, but  greater initial distances from the center of the Earth were considered. 
In computer simulations, bodies started from the point F located at a distance  $r_{ME}$  from the center of the Earth on the line from the Sun to the Earth.  
In different calculation variants, the distance $r_{ME}$ varied from $3r_E$  to $60r_E$, where $r_E$ is the radius of the Earth. 
The semi-major axis of the orbit of the Moon is equal to $60r_E$.
It is usually considered that the initial Moon formed close to the Earth, and then it increased its distance from the Earth.
The escape velocity for the Moon is equal to 2.38 km/s. 
In considered variants of calculations, initial velocities $v_{ej}$ varied from 2.5 to 16.4 km/s, and   values of $i_{ej}$  varied from 15$^\circ$ to 89$^\circ$. 
The gravitational influence of the Moon and 
 the velocity $v_M$ of the motion of the Moon around the Earth were not taken into account in calculations. For initial velocities of ejected bodies relative to the Sun, only the motion of the Earth around the Sun was added to ejection velocities. 
The vector of $v_M$ can have different directions as the Moon moves around the Earth. The value of $v_M$ is about 1 km/s for the present orbit of the Moon.The velocity of the lunar surface due to its rotation about the center of mass of the Moon is small and do not exceed 4 m/s.

At  $r_{ME}$=$5r_E$, the parabolic and circular velocities for the Earth are equal to 4.9  and 3.5 km/s, respectively. 
Note that the parabolic velocity is for eccentricity $e$ of the orbit equal to 1. Some ejected bodies can reach the surface of the Hill sphere of the Earth at $e$$<$1 and can leave this sphere even if their initial velocities are less than the parabolic velocity.
For ejection velocities less than the circular velocity for the Earth at a considered distance, all ejected bodies could fall onto the Earth or the Moon. 
As it is discussed in Section 2 and in (\cite{Ipatov2024}), the considered model does not calculate the probability of a collision of a body with the Moon before this body left the Hill sphere of the Earth for the first time.
The results of calculations obtained for  some values of $v_{ej}$ and $i_{ej}$ correspond to
the evolution of bodies for other values of real ejection velocities and inclinations. Such difference is smaller for greater $v_{ej}$.
In our studies we were interested in analyzing different types of evolution of orbits of bodies ejected from the Moon, but we were not interested in understanding the exact dependence of the evolution on initial data.

At $v_{ej}$=2.5 km/s and $r_{ME}$=$3 r_E$, the dynamical lifetime of all ejected bodies was less than 5 days. 
For $r_{ME}$=5$r_E$ and $v_{ej}$=2.5 km/s, some of the bodies left the Hill sphere of the Earth at 15$^\circ$$\le$$i_{ej}$$\le$$30^\circ$, though at greater   $i_{ej}$  most of the ejected bodies quickly fell onto the Earth. At $v_{ej}$=5 km/s, $T$=$T_{end}$, and $r_{ME}$=$5r_E$, the value of $p_E$ was about 0.3 for the considered ejection angles $i_{ej}$ from 15$^\circ$ to 89$^\circ$.
Note that the parabolic velocity for the Earth at $r_{ME}$=$5r_E$ is a little less than 5 km/s.
In some variants (e.g., at $r_{ME}$=$3r_E$, $v_{ej}$=5 km/s, $i_{ej}$$=$$30^\circ$; 
$r_{ME}$=$5r_E$, $v_{ej}$=2.5 km/s, 15$^\circ$$\le$$i_{ej}$$\le$$30^\circ$; 
$r_{ME}$=$7r_E$, $v_{ej}$=2.5 km/s, 15$^\circ$$\le$$i_{ej}$$\le$$45^\circ$) some ejected bodies could still move around the Earth after 100 Myr. For the model with the Moon included in integration, most of such bodies should collide with the Moon.

For the ejection of bodies from the present orbit of the Moon, $T$=$T_{end}$ and 15$^\circ$$\le$$i_{ej}$$\le$$89^\circ$, the value of $p_E$ was about 0.27–0.35, 
0.2-0.25, and 0.1-0.14  at $v_{ej}$=2.5 km/s,  $v_{ej}$=5 km/s, and  at 12$\le$$v_{ej}$$\le$16.4 km/s, respectively. 
The probabilities of collisions of bodies ejected from the Moon with the terrestrial planets were similar to those ejected from the Earth, but for different ejection velocities.

\section{Conclusions}
This study analyzes the motion of bodies ejected from the Earth or the Moon. We studied the ejection of bodies from several points on the Earth's surface, as well as from the most far point of the Moon from the Sun. Different velocities and angles of ejection of bodies were considered. The dynamical lifetimes of bodies reached a few hundred million years. Over the entire considered time interval, the values of the probability of a collision of a body ejected from the Earth with the Earth were approximately 0.3, 0.2, and 0.15–0.2 at $v_{ej}$ equaled to 11.5, 12, and 14 km/s, respectively. At $v_{ej}$$\le$11.3 km/s, most of the ejected bodies fell back onto the Earth. 
The total number of bodies delivered to the Earth and Venus probably did not differ much. The probabilities of collisions of bodies with Mercury and Mars usually did not exceed 0.1 and 0.02, respectively. 
At $v_{ej}$$ \ge$11.5 km/s, the probability of a collision of a body ejected from the Earth  with the Moon was about 15–35 times less than that with the Earth, and it was about 0.01. The probability of a collision with the Earth for a body ejected from the Moon moving in its present orbit was about 0.3–0.32,   0.2–0.22, and 0.1–0.14 
at $v_{ej}$=2.5 km/s, $v_{ej}$=5 km/s,  and at 12$\le$$v_{ej}$$\le$16.4 km/s, respectively. 

\section{Acknowledgements}
This work was supported by the Ministry of Science and Higher Education of the Russian Federation, within the budget theme of the Vernadsky Institute of Geochemistry and Analytical Chemistry of Russian Academy of Sciences.
I thank the  reviewer for helpful comments and suggestions.

\end{document}